\documentclass[english,copyright,creativecommons]{eptcs}
\usepackage[T1]{fontenc}
\usepackage[latin9]{inputenc}
\usepackage{wrapfig}
\usepackage{calc}
\usepackage{url}
\usepackage{amstext}
\usepackage{graphicx}

\makeatletter
\usepackage{tlatex}
\usepackage{color}
\definecolor{boxshade}{gray}{0.90}
\setboolean{shading}{true}

\providecommand{\event}{Formal Integrated Development Environment 2019 (F-IDE 2019)}
\def\authorrunning{M.A. Kuppe, L. Lamport \& D. Ricketts}
\def\titlerunning{The TLA+ Toolbox}
\author{Markus Alexander Kuppe
 \institute{Microsoft Research}
 \email{makuppe@microsoft.com}
 \and Leslie Lamport
 \institute{Microsoft Research}
 \and Daniel Ricketts
 \institute{Oracle Corporation}
 \email{dan.s.ricketts@gmail.com}
}
\newcommand{\tlaplus}{TLA$^{+}$}

\usepackage{tikz}
\usepackage{pgf-umlsd}
\usepgflibrary{arrows} 

\renewcommand{\textemdash}{---}

\@ifundefined{showcaptionsetup}{}{%
 \PassOptionsToPackage{caption=false}{subfig}}
\usepackage{subfig}
\makeatother

\usepackage{babel}
\usepackage{listings}

\begin{document}

\title{The \tlaplus{} Toolbox}

\maketitle

\begin{abstract}
We discuss the workflows supported by the \tlaplus{} Toolbox to write
and verify specifications. We focus on features that are useful in
industry because its users are primarily engineers. Two features are
novel in the scope of formal IDEs: \emph{CloudTLC} connects the Toolbox
with cloud computing to scale up model checking. A \emph{Profiler}
helps to debug inefficient expressions and to pinpoint the source
of state space explosion. For those who wish to contribute to the
Toolbox or learn from its flaws, we present its technical architecture.
\end{abstract}

\section{Introduction}

The \tlaplus{}\emph{ }Toolbox combines a set of (command-line) tools
into an integrated development environment (IDE). Its primary goal
it to make \tlaplus{} more widely used. To encourage the use of the
Toolbox, the Toolbox is made freely available. It is developed by
an open-source community.\footnote{\url{https://github.com/tlaplus/tlaplus}}
Section~\ref{sec:Toolbox-Features} describes the main Toolbox workflows
by showing how the $Simple$ spec is model-checked and mechanically
verified inside the Toolbox.\footnote{\url{https://github.com/tlaplus/Examples/tree/master/specifications/TeachingConcurrency}}
Section~\ref{Arch:Toolbox} shifts the discussion to the architecture
and test infrastructure of the Toolbox. We conclude with outlining
future engineering and research-related work.

\subsection{Background}

\tlaplus{} is a high-level, math-based, formal specification language
used at companies such as Amazon and Microsoft to design, specify,
and document systems \cite{Newcombe2015}. A system is specified by
formulas expressed in the Temporal Logic of Actions \cite{Lamport1994,Lamport2003},
a variant of Pnueli's original linear-time temporal logic \cite{Pnueli1977}.
Data structures are represented with Zermelo-Fränkel set theory with
choice. \tlaplus{} is an untyped language and thus simpler and more
expressive than programming languages \cite{Lamport1999}. Generating
code is not in the scope of \tlaplus{}. It is implementation language
agnostic and meant to find bugs above the code level. Further material
about \tlaplus{} is available \cite{Lamport2003,Lamport2009,Lamport2012,Lamport2019},
but is not needed for what follows.

The underlying tools, with which to check and reason about \tlaplus{}
specs, are the explicit state model checker TLC and the \tlaplus{}
proof system (TLAPS). While TLC is used to check a finite model of
a spec, TLAPS supports deductive reasoning about a spec with infinitely
many reachable states \cite{Chaudhuri2008,Yu1999}. Other \tlaplus{}
tools are a parser to syntactically and semantically check a spec
(SANY), a translator to transpile a PlusCal algorithm into \tlaplus{},
and a pretty-printer to render \tlaplus{} with \LaTeX.

\section{Toolbox Features\label{sec:Toolbox-Features}}

The main parts of the Toolbox are the Spec Explorer, spec editors,
model editors, and a Trace Explorer. The goal is to keep the interface
simple and intuitive. Functionality not needed by beginning users
is initially hidden.

\subsection{Spec Explorer\label{UI:Spec-Explorer}}

\begin{wrapfigure}{r}{0.3\columnwidth}%
\centering{}\includegraphics[scale=0.3]{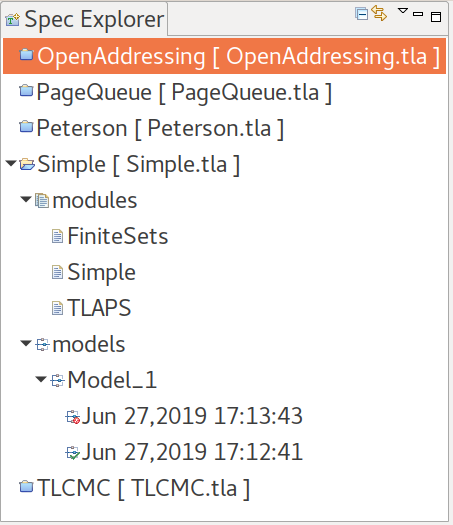}\caption{\label{fig:Spec-Explorer}Spec Explorer showing five \tlaplus{} specifications.
The active Simple specification consists of three modules and a single
model with two history items.}
\vspace{-0.75in}
\end{wrapfigure}%
The Toolbox can host multiple \tlaplus{} specifications. A spec has
a root module, a set of additional modules and \textemdash{} optionally
\textemdash{} a collection of models. All specifications are accessible
from the Spec Explorer (figure~\ref{fig:Spec-Explorer}), but users
can work on only one spec at a time. Contrary to most Eclipse-based
IDEs, specs can be exported from and imported into the Toolbox from
the file system.

\subsection{Spec Editor\label{UI:Spec-Editor}}

A spec editor provides the usual commands that are helpful to create
and edit \tlaplus{} modules. The editor can also generate and show
a pretty-printed version of the module. It shows annotations when
SANY finds parsing errors.

\subsubsection{PlusCal\label{UI:PlusCal}}

The PlusCal algorithm language is a formally defined and verifiable
pseudocode that is translated into \tlaplus{} \cite{Lamport2009}.
PlusCal resembles an imperative programming language and is especially
well suited to express sequential and shared-memory multithreaded
algorithms. Neither the \tlaplus{} concept of refinement nor fine-grained
fairness constraints can be expressed in PlusCal.

A PlusCal algorithm and its \tlaplus{} translation go in the same
\tlaplus{} module (figure~\ref{fig:TLAPS}). The editor has a command
to automatically translate a PlusCal algorithm into \tlaplus{}. Users
can quickly navigate to and from its \tlaplus{} translation, which
can help them understand the PlusCal code. Note that PlusCal is designed
to make the \tlaplus{} translation easy to read. The exposure to
\tlaplus{} is intended to encourage users to learn \tlaplus{}. Also,
as mentioned above, refinement and some fairness constraints can be
expressed only in the \tlaplus{} translation.

Combining a PlusCal algorithm and its \tlaplus{} translation into
a single spec poses the risk that they get out of sync unnoticed.
To alleviate this problem, a TLC feature of the next Toolbox release
will issue a warning if a translation has become stale.\footnote{\url{https://github.com/tlaplus/tlaplus/issues/296}}

\begin{figure}
\begin{centering}
\includegraphics[scale=0.4]{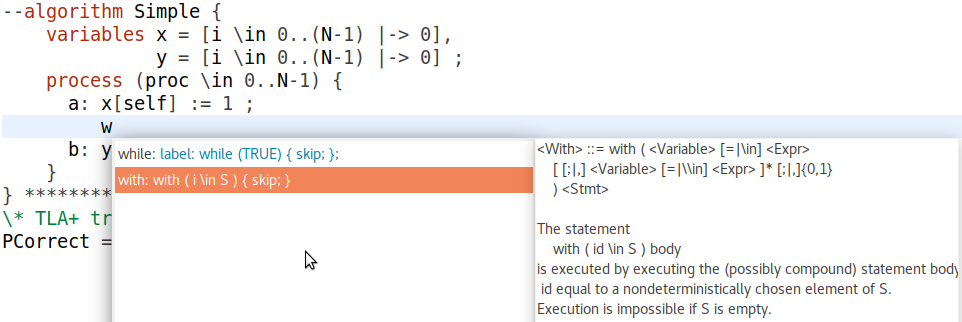}\caption{\label{fig:PlusCal-templates}Completion for PlusCal expressions.
The syntax and semantics of expressions are shown as well.}
\par\end{centering}
\end{figure}

As seen in figure \ref{fig:PlusCal-templates}, a spec editor provides
templates for PlusCal expressions.
 The goal of templates is not to speed up typing \textemdash{} specs
are usually short \textemdash{} but to guide novice users by putting
the syntactical and semantic documentation of PlusCal expressions
at their fingertips. The user experience also resembles that of familiar
programming IDEs, but the lack of types and type inference limits
word completion to the syntax level.

\subsubsection{Proof{\small{}s}\label{UI:TLAPS}}

\tlaplus{} proofs are not written interactively but are constructed
hierarchically \cite{Lamport1995}. Hierarchical proofs enable a non-linear
proof style: Individual steps of a proof can be checked independently
of other steps. Also, steps can be decomposed into simpler steps should
TLAPS fail to prove the higher-level steps.

\begin{figure}
\centering{}%
\fbox{\includegraphics[scale=0.245]{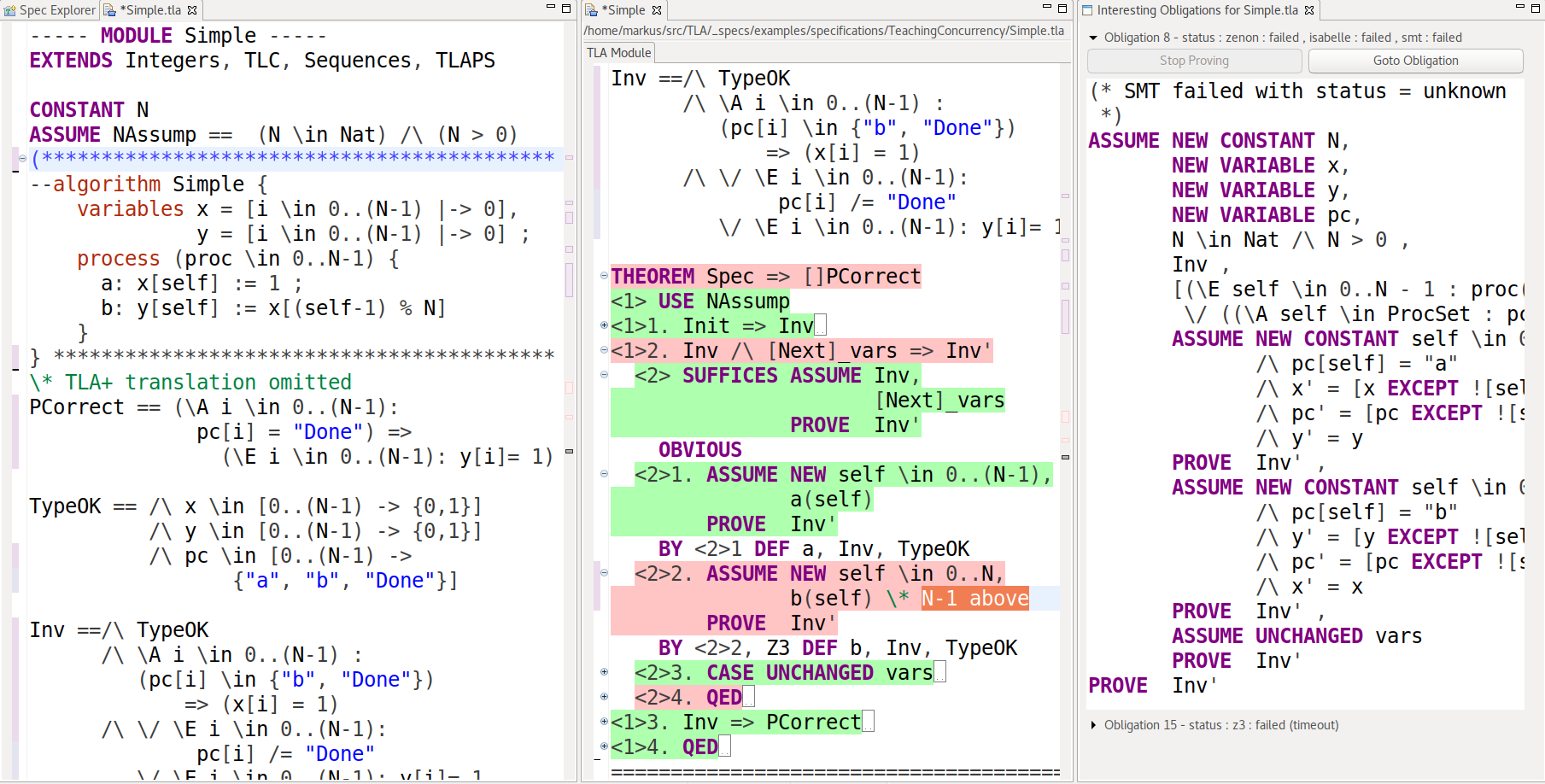}}\caption{\label{fig:TLAPS}A (split) spec editor showing a PlusCal algorithm
and related safety properties (left) as well as its partially collapsed
TLAPS proof (center). The <2>2 leaf step failed to prove (red). The
right-hand side displays the expanded hypothesis of the failed obligation.}
\end{figure}

In figure \ref{fig:TLAPS} we see a hierarchical \tlaplus{} proof
where lower-level, proven steps have been collapsed. The green color
indicates steps that TLAPS proved successfully while red indicates
steps that it failed to prove. A failed step is additionally displayed
in the logging part, showing exactly what TLAPS was trying to prove
(see the right-hand side of figure~\ref{fig:TLAPS}).

A wizard exists to decompose a step to be proved into a sequence of
simpler steps, based on the logical structure of the step. For example,
a step asserting the conjunction of properties can be proved by steps
that each assert one of those properties. The wizard can add these
steps as a lower level proof, and they can be verified separately
with TLAPS. \cite{Cousineau2012a}. 

TLAPS maintains a fingerprint of every proof obligation so that it
does not need to re-prove previously verified obligations. A user
tells TLAPS to forget fingerprints from the Toolbox.

Unlike TLC, which is part of the Toolbox, users have to manually install
TLAPS\@. The Toolbox will automatically find the TLAPS installation.

\subsection{Model \label{UI:Model}\label{Model-Editor}}

TLAPS can verify specifications with an infinite set of possible states.
The explicit-state model checker, however, would never finish if a
user tries to verify a spec with an infinite state space. A primary
goal of TLC is that the spec need not be modified to be checked. Instead,
a TLC model restricts the spec's state space to a finite set of possible
states by assigning concrete values to parameters and by stating additional
bounds. To gain higher confidence, users typically check a spec on
many separate models.

The Toolbox represents a TLC model by a model part, which also contains
settings for how TLC should be executed. Seasoned users will tune
these settings and define additional checks that TLC should perform.

Contrary to the proof system (section~\ref{UI:Spec-Editor}), the
Toolbox provides a structural representation of a model by grouping
related inputs into tabs and sections. To guide users, inputs are
validated when entered. Warnings and suggested fixes are placed next
to the inputs. The model is kept in sync with the spec so that relevant
changes, for example the addition or removal of a constant, are automatically
reflected in the model.

A model is associated with its respective spec in the Spec Explorer.
The Spec Explorer also maintains a history of model checker runs,
each history entry storing the model checking result and snapshots
of the spec's modules (figure~\ref{fig:Spec-Explorer} and section~\ref{UI:Result-Page}).
A user can compare modules, which for example will remind her of the
differences between long model checking runs.

A model is stored as an XML file in the file system, so it can be
checked into a source code management system. This way, the user can
reproduce model checking runs.

\subsubsection{CloudTLC\label{UI:CloudTLC}}

With CloudTLC users can shift from a primarily sequential model checking
workflow to one where any number of models can be checked concurrently.
Additionally, CloudTLC scales up explicit-state model checking to
more powerful hardware. This way, design variants or competing optimizations
can be quickly explored in parallel. This can be used either to reduce
overall specification time or to enable a broader exploration of the
design space.

CloudTLC provisions a set of cloud instances and runs TLC on them.
If a user opted to run model checking on multiple instances, CloudTLC
starts TLC in distributed mode \cite{Kuppe2014}. Upon completion
of model checking, CloudTLC instances terminate automatically after
a grace period. If the user keeps the Toolbox open, it will show model
checking progress and final results. Alternatively, the Toolbox may
be disconnected at any time; CloudTLC sends the results to a user-provided
email address in a format that can be imported into the Toolbox. In
other words, CloudTLC is completely transparent: Toolbox features
such as the Profiler or Trace Explorer (sections~\ref{UI:Profiler}
and \ref{UI:Trace-Explorer} below) work as if model checking runs
locally.

CloudTLC's startup time is dominated by the time it takes an IaaS
provider to spin-up instances. To reduce startup time, subsequent
model checker runs re-use previously provisioned instances unless
they have already terminated. CloudTLC can also be started by running
the Toolbox in command-line mode. This is useful for automation such
as running the TLC performance test suite (section~\ref{Arch:Testing}).

\subsubsection{Results\label{UI:Result-Page}}

The status and progress of TLC checking a model is shown on the results
part. The results include the start and end wall-clock time, the probability
of an incomplete state space exploration due to distinct states falsely
being considered equivalent, as well as global state-graph and action-level
statistics. State-graph statistics are:
\begin{itemize}
\item $Diameter$: The depth that TLC has reached in its exploration of
the state-graph
\item $Distinct\,States$: The cardinality of the set of reachable vertices
of the state-graph
\item $\left(Total\right)\,States$: The number of states examined by TLC
as part of model checking
\end{itemize}
The ratio of distinct to examined states approximates the degree of
the state-graph. Action statistics \textemdash{} further discussed
in section \ref{UI:Profiler} \textemdash{} are similar to global
state-graph statistics, except that they are reported at the level
of individual \tlaplus{} actions and the diameter is undefined.

Plots of some statistics can be shown that indicate the remaining
model checking time.\footnote{The cardinality of the set of unexplored states plotted over time
looks like a parabola.} Furthermore, a state-graph may be visualized graphically. However,
this is possible only for small graphs due to the inherent complexity
of laying out directed graphs. The description of a graph can be exported
to specialized tools such as Cytoscape \cite{Shannon2003}.

\subsubsection{Profiler\label{UI:Profiler}}

Engineers primarily rely on model checking, not theorem proving,
and they want to verify large models. This makes them not only users
of CloudTLC (section~\ref{UI:CloudTLC}), but also of the Toolbox's
Profiler. Profiling a \tlaplus{} model and its spec produces four
different types of data:
\begin{itemize}
\item $Invocation~count$: The number of times a \tlaplus{} expression
is evaluated
\item $Cost$: The number of operations performed to enumerate the elements
of a set or sets, should enumeration be required to evaluate an expression
\item $States/Action$: The number of states a particular \tlaplus{} action
caused to be examined
\item $Distinct~States/Action$: The number of distinct states an action
produced
\end{itemize}
\begin{figure}
\centering{}\subfloat[\label{fig:ProfilerA}]{\begin{centering}
\includegraphics[scale=0.225]{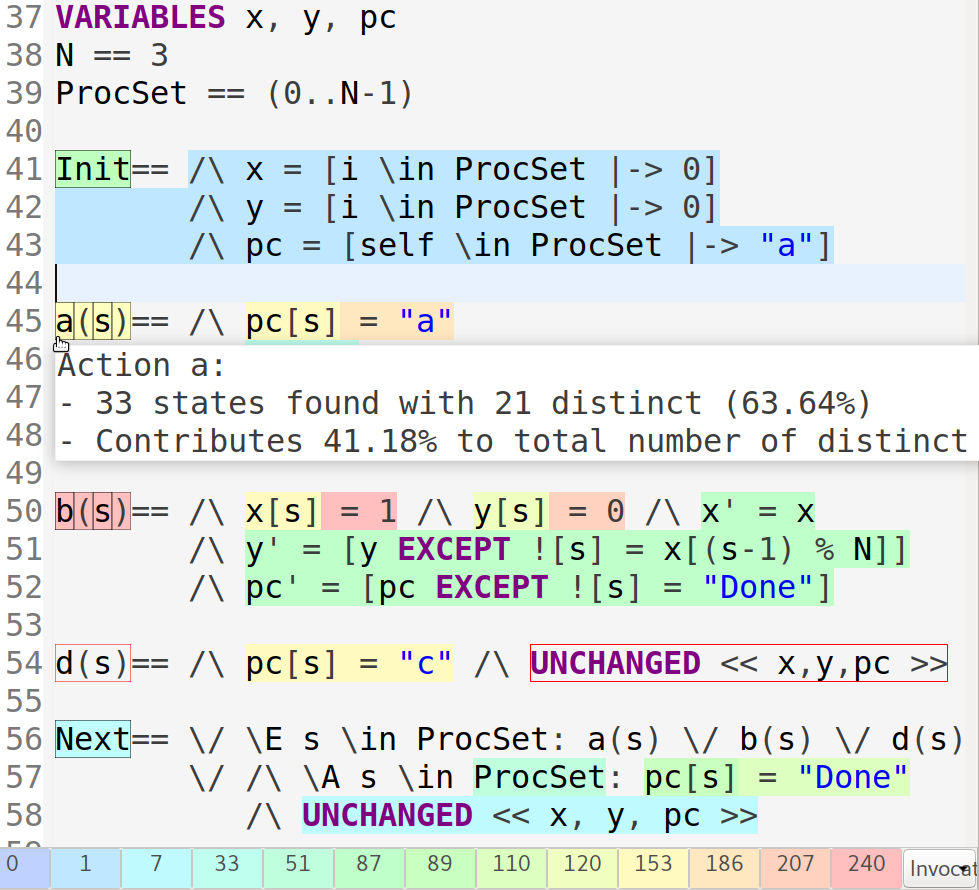}
\par\end{centering}
}\subfloat[\label{fig:ProfilerB}]{\begin{centering}
\includegraphics[scale=0.225]{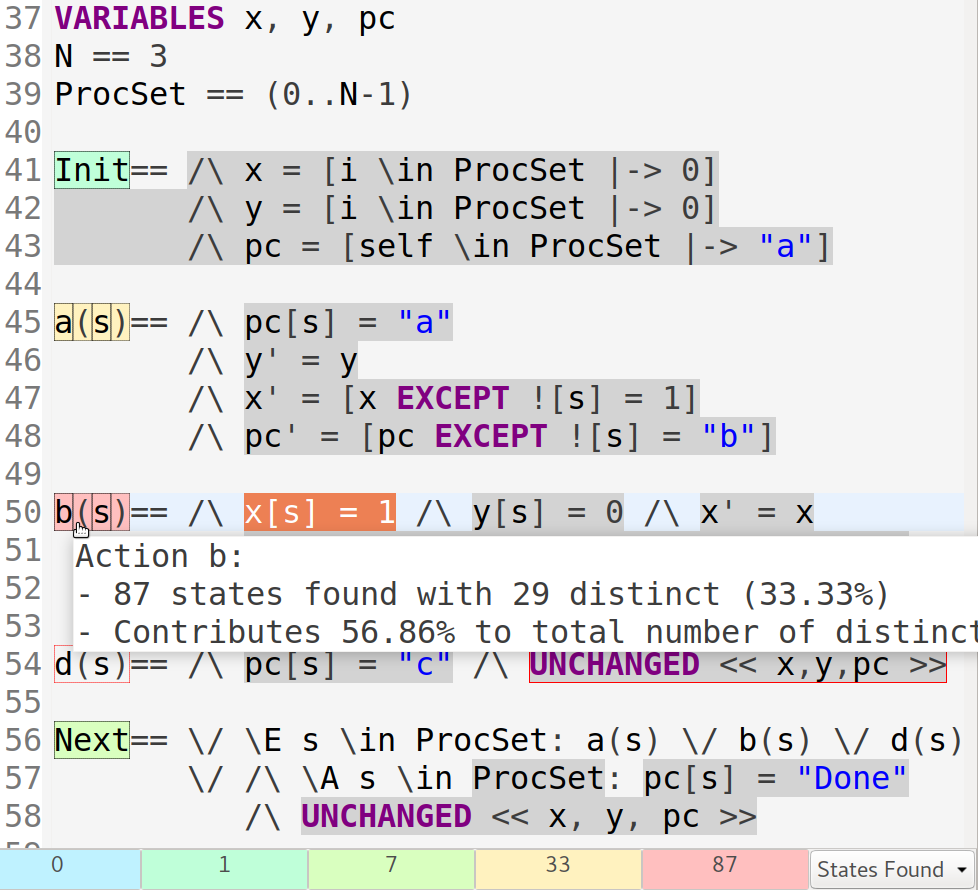}
\par\end{centering}
}\caption{\label{fig:Profiler}The editor highlights individual action and evaluation
statistics. Red boxes indicate actions that are never enabled and
dead expressions. The hover help displays \tlaplus{} action statistics.
Clicking the heatmap at the bottom selects the corresponding expression
or action.}
\end{figure}
We call the first two data types evaluation statistics and collect
them at the global as well as at the call-chain level. To explain
their differences, we assume an identical, constant cost for all expressions.
This way, a user can identify the biggest contributor to overall model
checking time simply by looking at the number of invocations. However,
some expressions require the model checker to explicitly enumerate
data structures for which costs are the quantitative measure: For
example, let $S$ be the set of natural numbers from $N$ to $M$
such that $N\ll M$ in the \tlaplus{} expression $\forall s\in\textsc{{subset}}\,S:s\subseteq S$.
The cost of evaluating the expression is determined by the number
of operations required by TLC to enumerate all the subsets in $\textsc{{subset}}\,S$
(the powerset of $S$). This expression will be a major contributor
to model checking time even if its number of invocations is low.

With action statistics, a user can pinpoint the source of state space
explosion at the action level. To explain their application, consider
figure \ref{fig:ProfilerA}, which is a deliberately inefficient \tlaplus{}
translation of the $Simple$ algorithm shown in figure \ref{fig:TLAPS}.
The enablement predicate on line 50 of action $b$, consisting of
the two expressions $x\left[s\right]=1$ and $y\left[s\right]=0$,
has been globally invoked the most. Related, action $b$ has been
found to produce 87 states of which only 29 are distinct (see hover
help in figure~\ref{fig:ProfilerB}). In contrast, action $a$ is
much more efficient in terms of its total states to distinct states
ratio. This draws the user's attention to the enablement predicate
of action $b$, which is weaker compared to the predicate $pc\left[s\right]=``a"$
of action $a$. The enablement predicate of $b$ is true of states
for any value of $pc$. While this does not violate the specification's
safety properties, changing the enablement predicate of $b$ to $pc\left[s\right]=\mbox{``b''}$
puts the ratio of total to distinct states in the region of the same
ratio of action $a$. It also corresponds to the regular translation
of the $Simple$ algorithm. Additionally, the enablement of action
$d$ on line 54 is evaluated 153 times. Yet, the numbers of total
and distinct states are zero as indicated by the red boxes. This suggests
that there is either an error in the spec, or else that the user can
delete action $d$.

As shown in figure \ref{fig:ProfilerA} and \ref{fig:ProfilerB},
the Toolbox highlights expressions with colors chosen from a one-dimensional
heatmap based on an expressions' corresponding evaluation statistic.
Expressions with the highest number of invocations will be highlighted
with a red color whereas those with no invocations will be dark blue.
The heatmap appears as a legend at the bottom of the editor and reveals
the corresponding editor location when clicked. Users can switch what
the highlighting is based on between invocation counts, costs, and
the number of examined or distinct states. Additionally, users focus
on a single call-chain when they select an expression in the editor.

In summary, the Profiler makes different kinds of inefficiencies explicit.
For those inefficiencies related to the evaluation of expressions,
the model provides a way for users to override \tlaplus{} operators
with more efficient variants. An even more extreme optimization is
to override \tlaplus{} operators with equivalent Java functions.
Action statistics expose weaknesses in enablement predicates or spec
errors.

The Profiler requires neither the specification nor the model to be
modified to collect statistics. However, profiling has a non-negligible
performance overhead and should be disabled when checking large models.
Profiling is unavailable when TLC runs in distributed mode.

\pagebreak{}

\subsubsection{Trace Explorer\label{UI:Trace-Explorer}}

\begin{wrapfigure}{R}{0.5\columnwidth}%
\centering{}%
\fbox{\includegraphics[scale=0.33]{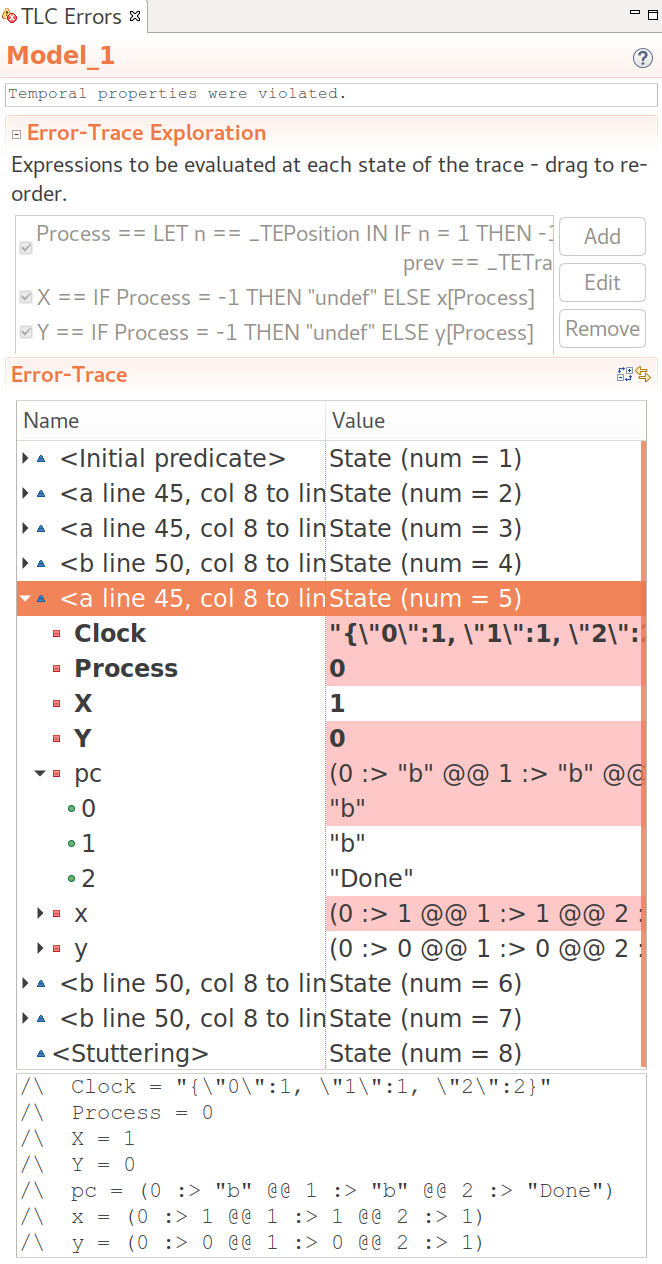}}\caption{\label{fig:TraceExplorer}A trace of a violated liveness property
(trace expressions in bold). The definitions of three trace expressions
are shown above the trace. The expanded state shows the values of
all variables and trace expressions for state number five.}
\vspace{-0.25in}
\end{wrapfigure}%
Should model checking find a violation of any of the stated safety
or liveness properties, the corresponding error trace will be displayed
in the Trace Explorer as shown in figure \ref{fig:TraceExplorer}.
An error trace is a sequence of states, and a state is an assignment
of values to variables.

The Trace Explorer colors the variables of a state that have changed
from the predecessor state. Navigating from a state in the trace to
the location in the editor of the corresponding \tlaplus{} action
(and if applicable, the corresponing PlusCal expression) is also supported.
The Trace Explorer can handle traces with thousands of states.

To study traces, the Trace Explorer supports the evaluation of trace
expressions. A trace expression can be a function of a state or a
pair of successive states \cite[p. 4-5]{Lamport1994}. A trace expression
may optionally be named to facilitate expression composition. A trace
expression may be built from all operators in the scope of the root
module. In addition, two built-in operators are available:
\begin{description}
\item [{\_TEPosition}] Equal to the position of the corresponding state
in the error trace
\item [{\_TETrace}] A \tlaplus{} sequence of states such that \texttt{\_TETrace{[}\_TEPosition{]}}
equals the state at position \texttt{\_TEPosition} in the error trace
\end{description}
Note that these operators increase the expressiveness of trace expressions
such that they can be formed from the collection of variables of all
states of the trace. For example, trace expressions can compare the
values of variables of two or more arbitrary states (see trace variables
\texttt{Clock, Process, X, }and\texttt{ Y} in figure~\ref{fig:TraceExplorer}
and listing~\ref{fig:backend-output}). Users can write such expressions
to format traces that can be directly copied to third-party tools
\cite{Kuppe2019}.  Internally, the Toolbox evaluates trace expressions
by generating a special \tlaplus{} module that is checked by using
the functionality discussed in section~\ref{Arch:Back-end-integration}.

\section{Toolbox Architecture\label{Arch:Toolbox}}

We are trying to make the Toolbox an industrial strength integrated
development environment with support for all aspects of writing \tlaplus{}
specs. If possible, features are implemented at the back-end layer
to satisfy command-line aficionados, to support re-use, and to integrate
with automation. Features not core to \tlaplus{}, such as visualizations
of the state-graph and error traces, are left to specialized third-party
tools to which relevant information is exported in compatible data
formats.

To attract a wide range of users, the Toolbox is compatible with the
three most common operating systems: macOS, Windows, and Linux. The
Toolbox requires no external dependencies except to generate PDFs.
Primarily, this is because the Toolbox is written in Java and built
on the Eclipse Rich Client Platform (RCP) \cite{McAffer2010}. \label{Arch:Eclipse}RCP
provides a number of IDE features such as a help system, a desktop
notification system, and an update manager. RCP also defines usability
guidelines and best practices that, while not always applicable to
a formal integrated development environment, help enforce a consistent
user experience. Others can add functionality by contributing extensions
and OSGi services \cite{Marples2001}.

Building the Toolbox on top of RCP is not without drawbacks. RCP follows
a quarterly release schedule. Upgrading the Toolbox to a new RCP release
frequently leads to subtle bugs that are not detected by our tests.
The rate at which macOS, Windows, and Linux innovate forces us to
upgrade. This causes significant manual work. 

The Spec Explorer (section~\ref{UI:Spec-Explorer}) is the primary
interface to work with specs. However, users can also create, move,
or modify files at the OS level. This feature causes many incompatibilities
and bugs, since RCP wants files to be modified through its file system
abstraction. As a result, the Spec Explorer and editors get out of
sync with the actual files. In hindsight, file operations should have
been restricted to the Toolbox or proper support should have been
added to RPC; the workarounds at the Toolbox layer continue to cause
bugs.

Throughout the development of the Toolbox, it has become clear that
RCP has to be considered a whitebox component. In other words, projects
built on top of RCP have to accept the burden of co-ownership and
contribute to its development.

\subsection{Back-end integration\label{Arch:Back-end-integration}}

For latency reasons, the short-lived lightweight \tlaplus{} tools
such as the SANY parser, the PlusCal translator, and the pretty-printer
are executed as part of the Toolbox process. Extending this functionality
is possible only via extensions and OSGi services, as discussed in
section \ref{Arch:Toolbox}. However, the heavyweight model checker
and proof system spawn as separate processes. There are three reasons
to spawn separate, per-invocation processes:
\begin{itemize}
\item A back-end cannot be executed in-process because it is implemented
in a language that cannot execute on the Java VM\@. This is true
for TLAPS, which is written in OCaml. 
\item Process separation acts as a circuit breaker; a crash of either a
back-end or the Toolbox does not interfere with the other process.
For example, we do not want a Toolbox crash to also crash a long-running
model checker run. This safeguard is especially important while a
back-end matures.
\item The Java VM's runtime parameters are fixed after Toolbox startup.
Running a back-end as part of the Toolbox causes the back-end to inherit
the Toolbox's parameters. The resource requirements of back-ends are
usually different from those of the Toolbox \textemdash{} for example,
model checkers have very high resource requirements (section~\ref{UI:CloudTLC}).
\end{itemize}

\subsubsection{Toolbox to Back-end}

The Toolbox relies on an RCP framework to provide user-visible progress
reports and cancellation support for the back-ends that it spawns.\footnote{\url{https://www.eclipse.org/articles/Article-Concurrency/jobs-api.html}}
For that, back-ends have to implement an adapter that sets command-line
parameters. The primary parameters include the path to the \tlaplus{}
spec. In addition, parameters may include performance-specific settings
such as the number of cores a back-end may use.

\label{Arch:DisadvantageTLCTextBasedAPI-1}The Toolbox serializes
a subset of the model into a plain-text configuration file (section~\ref{UI:Model}).
This file contains the \tlaplus{} behavior spec as well as the invariants
and properties to be checked. It also includes definitions for all
declared constants. Optionally, it may contain definition and operator
overrides as well as state and action constraints. The configuration
is not specific to TLC and is therefore reusable by other back-ends.

\subsubsection{Back-end to Toolbox}

\label{Arch:TextBasedBackEndAPI}The Toolbox parses back-end progress
and results with a framework that is connected to the Toolbox UI via
the Model-View-Presenter (MVP) design pattern.\footnote{\url{https://en.wikipedia.org/wiki/Model-view-presenter}}

\begin{lstlisting}[caption={A chunk of TLC back-end output containing
a single multi-line statement (line 4 to 11). The statement encodes
a state (constant 2217:4) and corresponds to the expanded state shown
in figure \ref{fig:TraceExplorer}.},label={fig:backend-output},numbers=right,numberstyle={\scriptsize},basicstyle={\footnotesize\ttfamily},frame=top,frame=bottom,captionpos=b]
" @@ 3 :> "a" @@ 4 :> "a")
@!@!@ENDMSG 2217 @!@!@
@!@!@STARTMSG 2217:4 @!@!@
5: <next_action line 175, col 3 to line 209, col 2 of module TE>
/\ X = 1
/\ Y = 0
/\ Process = 2
/\ Clock = "{\"0\":1, \"1\":2, \"2\":1}"
/\ x = (0 :> 1 @@ 1 :> 1 @@ 2 :> 1)
/\ y = (0 :> 0 @@ 1 :> 1 @@ 2 :> 0)
/\ pc = (0 :> "b" @@ 1 :> "Done" @@ 2 :> "b")
@!@!@ENDMSG 2217 @!@!@
@!@!@STARTMSG 2217:4 @!@!@
6: <next_action line 220, col 3 to line 254, col 2 of module TE>
/\ X = 
\end{lstlisting}
A back-end specific parser has to read a stream of back-end output
that can consist of variably sized chunks of incomplete print statements.
For efficiency reasons \textemdash{} error traces occasionally contain
thousands of states \textemdash{} and to simplify the implementation,
parsing is based on special tokens and constants that wrap the output.
Parsing is performed in three stages: 
\begin{enumerate}
\item It buffers chunks of characters into lines separated by a newline
character
\item It identifies the lines that belong to a multi-line statement with
the help of special tokens
\item It de-serializes a multi-line statement into Java objects via MVP
\end{enumerate}
Listing \ref{fig:backend-output} shows an example of a multi-line
statement. In some cases, such as when error traces are serialized,
the multi-line statement is partially formatted to be valid \tlaplus{}.

In summary, the back-end integration is at a low enough level to free
back-ends from providing high-level interfaces such as a REST API\@.
Instead, a back-end's existing text-based IO can be reused. The flexibility
of this low-level integration comes at a price: the lack of compile-time
validation makes the evolution of back-ends more difficult. For the
model checker, this problem is alleviated by following a synchronized
release schedule. However, the evolution is still error-prone and
requires significant testing of the back-end and the Toolbox (section~\ref{Arch:Testing}).
Performance problems, related to parsing large outputs, require low-level
implementation optimizations. The maturity of the TLAPS back-end means
that its inputs and outputs seldom change. The Toolbox does not support
graceful back-end termination because the integration makes this impossible.
\label{JMX}Instead, the Toolbox could control TLC via its existing
Java Management Extension \cite{Perry2002}.

\subsubsection{CloudTLC Back-end\label{Arch:CloudTLC}}

\begin{wrapfigure}{R}{0.55\columnwidth}
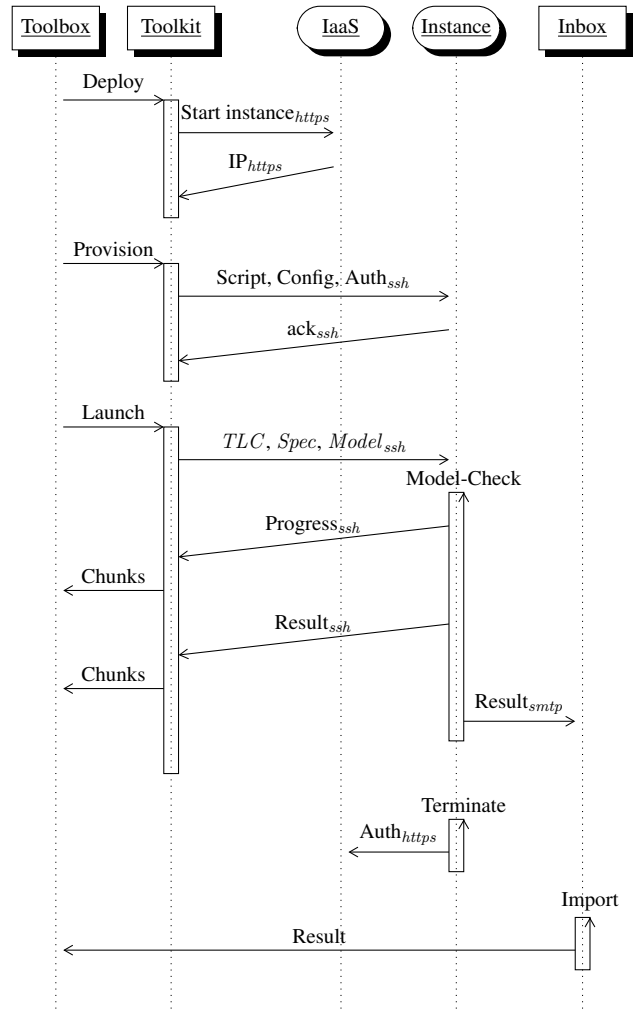
%
\centering
\tikzset{
  every picture/.append style={
    transform shape,
    scale=.725
  }
}

\begin{sequencediagram}

\newinst{tb}{Toolbox}
\newinst[.5]{tk}{Toolkit}

\tikzstyle{inststyle}+=[rounded corners=3mm]
\newinst[1.5]{iaas}{IaaS}
\newinst[.5]{inst}{Instance}

\tikzstyle{inststyle}+=[rounded corners=0mm]
\newinst[.7]{mail}{Inbox}


\begin{messcall}{tb}{Deploy}{tk}
  \mess{tk}{Start instance$_{https}$}{iaas}
  \mess[1]{iaas}{IP$_{https}$}{tk}
\end{messcall}

\begin{messcall}{tb}{Provision}{tk}
  \mess{tk}{Script, Config, Auth$_{ssh}$}{inst}
  \mess[1]{inst}{ack$_{ssh}$}{tk}
\end{messcall}

\begin{messcall}{tb}{Launch}{tk}
  \mess{tk}{$TLC$, $Spec$, $Model$$_{ssh}$}{inst}
    \begin{messcall}{inst}{Model-Check}{inst}
      \mess[1]{inst}{Progress$_{ssh}$}{tk}
      \mess{tk}{Chunks}{tb}
      \mess[1]{inst}{Result$_{ssh}$}{tk}
      \mess{tk}{Chunks}{tb}
      \mess{inst}{Result$_{smtp}$}{mail}
    \end{messcall}
\end{messcall}


\begin{messcall}{inst}{Terminate}{inst}
  \mess{inst}{Auth$_{https}$}{iaas}
\end{messcall}

\begin{messcall}{mail}{Import}{mail}
  \mess{mail}{Result}{tb}
\end{messcall}

\end{sequencediagram}\caption{\label{fig:UMLCloudTLC}Sequence diagram of the Toolbox running CloudTLC\@.
Message labels show the logical payload and subscripts indicate the
communication protocol. For simplicity, the deploy phase shows CloudTLC
executing with a single instance.}
\end{wrapfigure}%
CloudTLC has been implemented directly as part of the Toolbox (section~\ref{UI:CloudTLC}
above). CloudTLC is built on top of the frameworks discussed in section
\ref{Arch:Back-end-integration} and a multi-cloud toolkit that provides
an abstraction from individual IaaS providers.\footnote{\url{https://jclouds.apache.org/}}
The sequence diagram in figure \ref{fig:UMLCloudTLC} shows the interaction
between the building blocks of CloudTLC:
\begin{description}
\item [{Deploy}] The Toolbox queries the chosen IaaS provider via https
for specifically tagged CloudTLC instances. If the query result is
empty, the Toolkit requests the IaaS provider to launch the given
number of instances. If it is non-empty, the Toolbox starts the returned
instances and skips the next provisioning phase.
\item [{Provision}] The Toolbox configures the stock OS and installs dependencies
of the model checker. Authentication credentials, needed by later
phases of CloudTLC, are copied from the Toolbox's environment variables.
The provisioning phase directly communicates with each instance via
ssh.
\item [{Launch}] The Toolbox copies the spec and the model to the instance
and starts the model checker. The model checker continuously streams
its output to the local Toolbox. TLC also sends the result to a user-provided
email address.
\item [{Terminate}] The instance will wait for a grace period for subsequent
connection attempts. Afterwards, it terminates itself if and only
if email delivery has succeeded in the previous phase. Some IaaS providers
require the termination request to be authenticated, in which case
the previously mentioned credentials are used.
\end{description}
Because the hardware specifications of cloud instances are known during
development, the model checker can be optimally deployed and configured.
For example, the OS and TLC runtime parameters are hard-coded. CloudTLC
has been implemented for TLC, but it can be easily extended to new
back-ends. Likewise, it currently supports only Microsoft Azure, Amazon
AWS, and Packet Net. Support for other IaaS providers is possible
because of the multi-cloud toolkit. 

\subsection{Testing\label{Arch:Testing}}

The Toolbox development follows a combination of the test-driven and
the test-last methodologies. UI tests are written by Toolbox developers
and dedicated test engineers. A test suite of 178 unit, functional,
and end-to-end tests validates the main workflows of the Toolbox.
However, the tests do not cover TLAPS\@. Dedicated test suites exist
for the model checker and the proof system.

The Toolbox's test execution is fully automated and runs on macOS,
Linux, and Windows as part of the automated build. Builds are executed
by a continuous integration system. While test results do not get
published, the build output of the continuous integration system is
publicly available.\footnote{\url{https://nightly.tlapl.us/}} A test
suite of user-provided, real-world specifications continuously checks
the performance of TLC executed with CloudTLC\@. Overall, the test
suites provide a useful safety measure to catch functional and performance-related
regressions early in the development life-cycle.

\section{Conclusion}

Toolbox users write, model check, and deductively reason about \tlaplus{}
specs. The Toolbox has two features, which formal IDEs usually do
not provide: 
\begin{itemize}
\item CloudTLC connects the \tlaplus{} Toolbox with cloud computing to
enable users to check larger models and to explore the design space
faster
\item The Profiler assists users in identifying computationally expensive
\tlaplus{} expressions and to diagnose state space explosion
\end{itemize}
The discussion of its architecture and test infrastructure should
enable others to add new Toolbox features and will hopefully inspire
the development of more formal IDEs.

Whether or not the Toolbox can be considered successful in making
the \tlaplus{} specification language more widely used is difficult
to answer.\footnote{The Toolbox asks users to share TLC executions statistics and to identify
Toolbox installations.} Its previous release has been downloaded approximately 20k times
whereas the standalone model checker has seen one-tenth of this number.
The Toolbox has users at companies such as Amazon and Microsoft and
it has an active community of contributors.

\section{Future Work}

\paragraph{Proofs}

The main challenge in writing an invariance proof is finding the inductive
invariant. Trying to prove an invalid inductive invariant wastes a
lot of time. Model checking is useful to validate inductive invariant
candidates despite the enormous state space attached to it \cite{Lamport2018}.
The trick is to randomly select and check a subset of all type-correct
initial states. With this approach, repeated model checking finds
violations with high probability after only a few runs. However, the
Toolbox does not aggregate the results of repeated model checking.

Taming state space explosion by randomly selecting subsets of all
reachable states is also useful to validate for model checking lower-level
proof steps \cite{Lamport2019}. The Toolbox should provide help for
doing this.\vspace{-0.1in}

\paragraph{Profiler}

Engineers have expressed interest in overriding inefficient expressions
with Java functions. This is not easy to set up, especially for users
unfamiliar with Java. Besides the setup of overrides, it is an open
question how users can assert the equivalence of Java functions with
the \tlaplus{} expression it is overriding.\vspace{-0.1in}

\paragraph{Trace Exploration}

The Trace Explorer is good for finding the source of an error, but
its textural representation is not very good for understanding the
dynamics of a system. Schultz pioneered a graphical trace animator
that visualizes traces at the level of the problem domain \cite{Schultz2018}.
The layout is specified in \tlaplus{}, so users need not learn a
new visualization language. The next Toolbox release will incorporate
this graphical trace animator.\vspace{-0.1in}

\paragraph{Back-ends}

Konnov et al.\@ built a symbolic model checker for \tlaplus{}, the
results of which are encouraging \cite{Konnov2019a}. We wish to integrate
this model checker into the Toolbox as a new back-end. However, the
symbolic model checker requires users to provide type information
when its type inference fails. How the Toolbox can help users to work
with types is unclear. Note that the type inference of the symbolic
checker could also be used to provide a more powerful completion support
in the spec editor.

\section*{Acknowledgement}

The development of the \tlaplus{} Toolbox is a joint effort of INRIA
and Microsoft Research. Simon Zambrovski wrote and maintained the
initial version of the Toolbox. The reviewers provided useful comments
and suggestions.

\appendix
\bibliographystyle{eptcs}
\bibliography{bibtex}

\end{document}